\documentstyle[12pt]{article} 
\def\Box{\hbox{$\sqcup$\kern-0.66em\lower0.03ex\hbox{$\sqcap$}}}

\begin{document}

\begin{titlepage}
\begin{flushright}
IFUP--TH 38/96
\end{flushright}
\vskip 1truecm
\begin{center}
\Large\bf
Diffeomorphism invariant measure for  \\
finite dimensional geometries
\footnote{This work is  supported in part
  by M.U.R.S.T.}.
\end{center}

\vskip 1truecm
\begin{center}
{Pietro Menotti and Pier Paolo Peirano} \\ 
{\small\it Dipartimento di Fisica dell'Universit\`a, Pisa 56100, 
Italy and}\\
{\small\it INFN, Sezione di Pisa}\\
\end{center}
\vskip .8truecm
\begin{center}
July 1996
\end{center}
\end{titlepage} 

\begin{abstract}

We consider families of geometries of $D$--dimensional space,
described by a finite number of parameters. Starting from the De Witt
metric we extract a unique integration measure which turns out to be a
geometric invariant, i.e.\ independent of the gauge fixed metric used
for describing the geometries. The measure is also invariant in form
under an arbitrary change of parameters describing the geometries. We
prove the existence of geometries for which there are no related gauge
fixing surfaces orthogonal to the gauge fibers. The additional
functional integration on the conformal factor makes the measure
independent of the free parameter intervening in the De Witt
metric. The determinants appearing in the measure are mathematically
well defined even though technically difficult to compute.

\end{abstract}

\section{Introduction}
\label{introd}

The aim of discrete formulations of quantum gravity is to regularize
the theory by reducing the degrees of freedom to a finite number. The
underlying idea is to obtain the continuum theory by letting go to
infinity the number of degrees of freedom.

Quantum gravity in the functional approach is specified by an action
invariant under diffeomorphisms and an integration measure. If one
follows the analogy with gauge theories the analogous of the field
$A_\mu$ is the metric field $g_{\mu\nu}$ and the analogous of the gauge
invariant metric is the De Witt supermetric, which is the unique
ultra--local distance in the space of the metrics. The requirement of
ultra--locality i.e. the absence of derivatives in the metric is
dictated by the fact that the integration measure should not play a
dynamical role but only a kinematical one. While in dynamical
triangulations one replaces the functional integral by a discrete sum,
a typical example of the reduction to a finite number of degrees of
freedom is provided by Regge gravity \cite{catterall}. We shall
consider here a general situation in which the class of geometries
described by a finite number of parameters is not necessarily the
Regge model. 

In previous papers \cite{pmppp}, concerned with the $D=2$ case, the
breakdown to a finite number of degrees of freedom was achieved by
restricting the functional integral in the conformal gauge to those
conformal factors describing Regge surfaces.

This was possible due to the simplifying feature occurring in $D=2$
where all geometries can be described by the conformal factor and a
finite number of Teichm\"uller parameters. In $D>2$, to which we shall
address here, the scheme has to be enlarged.

Diffeomorphisms play a key role in the formulation of gravity and the
viewpoint we shall adopt is to treat them exactly at every stage. We
shall consider a class of geometries parameterized by a finite number
of invariants $l_{i}$ and described by a gauge fixed metric
$\bar{g}_{\mu\nu} (x,l)$. The functional integration will be performed
on the entire class $[f^{\star}\bar{g}_{\mu\nu}(l)](x)$ with $f$
denoting the diffeomorphisms \cite{jev}.  In other words the reduction
to a finite number of degrees of freedom will involve the geometries
only, not the diffeomorphisms. Since the integration on the latter is
infinite dimensional the related contribution will be a true
functional integral (the Faddeev--Popov determinant).

We recall that the differential structure of a manifold, i.e. the
charts and the transition functions, are to be given before imposing
on the differential manifold a metric structure. In other words if we
consider families of metrics on the same differential manifolds the
transition functions have to be independent of the metric
themselves. Such a feature is essential if we want that the variations
of the metric tensor appearing in the De Witt distance are to be
tensors under diffeomorphisms, or equivalently if the De Witt distance
has to be an invariant under diffeomorphisms.

In sect.2 after setting up the general framework we shall show that it
is possible to obtain from the De Witt supermetric a unique
integration measure given by a functional of
$\bar{g}_{\mu\nu}(x,l)$. This will be a geometric invariant that
remains unchanged also under arbitrary $l$--dependent diffeomorphisms.
Thus while the De Witt metric is invariant only under $l$--independent
diffeomorphisms the final expression of the associated measure can be
computed on charts with $l$--dependent transition functions.  In
addition the approach turns out invariant in form under an arbitrary
change of the parameters which describe our geometries.

Great simplifications occur if it is possible to choose a gauge fixing
surface in such a way that the variations of the metric under a
variation of the $l_{i}$ result orthogonal to the gauge fibers. On the
other hand we shall show that this can be realized only for special
classes of geometries. In different words only for selected
minisuperspaces this simplifying feature can be achieved.

Despite a gauge fixing procedure is necessary in order to factorize
the infinite volume of the diffeomorphisms we shall see that, provided
the chosen parameters $l_{i}$ are geometric invariants, no Gribov
phenomenon occurs.  It is well known that in the De Witt supermetric
an arbitrary parameter $C$ appears. Keeping the number of parameters
$l_{i}$ finite we shall show that in general such a dependence does
not disappear.  In sect.3 we enlarge the integration to the inclusion
of the conformal factor in addition to a finite number of parameters
$\tau_{i}$ which describe deformations transverse (i.e.\ non
collinear) to the orbits generated by the conformal and the
diffeomorphism groups. As a result of the functional integration on
the conformal factor the dependence on $C$ disappears. In $D=2$ the
relevant functional determinant is given by the exponential of the
Liouville action \cite{allc}. The analogous functional determinant in
$D>2$ is also perfectly defined, being the Lichnerowicz operator
elliptic in any dimension.  On the other hand the usual technique
which works in $D=2$, based on the local variation of the conformal
factor \cite{allc}, fails to work in $D>2$ due to the lack of
ellipticity of one of the operators entering in the conformal
variation.

\section{Geometric invariant measure}
\label{two}
In this paper we shall confine ourselves to Euclidean gravity, which
allows a positive definite De Witt supermetric.  We shall consider a
class of geometries parameterized by a finite number $N$ of parameters
which we shall call $l = \{l_i\}$. In the case of Regge geometry
one can think of the $l_i$ as the link lengths, but any other
parameterization is equally possible, as our treatment will be
invariant under the change of parameterization. For a given $l$ the
geometry is described by an infinite family of metrics, related
by diffeomorphism transformations. With $\bar g_{\mu\nu}(x,l)$ we
shall denote a special choice (gauge fixing) of the metric describing
the given geometries. The choice is widely arbitrary; we shall see
that the result will be independent of such a choice. The diffeomorphism
transformations act on $\bar g_{\mu\nu}(x,l)$ as follows
\begin{equation}
g_{\mu\nu}(x,l,f) = [f^{\star}\bar g_{\mu\nu}(l)](x) = \bar
g_{\mu'\nu'}(x'(x),l) 
{\partial x^{\mu'}\over \partial x^\mu}{\partial x^{\nu'}\over
\partial x^\nu}.
\end{equation}

As explained in the introduction we shall consider the metric
$g_{\mu\nu}(x)$ as the basic integration variable and as functional
integration measure we adopt the one induced by the De Witt
supermetric  \cite{dewitt}
\begin{equation}
\label{dewittm}
(\delta g,\delta g) = \int \sqrt{g(x)}\, d^D x \,\delta
g_{\mu\nu}(x)
G^{\mu\nu\mu'\nu'}(x) \delta g_{\mu'\nu'}(x)
\end{equation}
with
\begin{equation}
G^{\mu\nu\mu'\nu'} = g^{\mu\mu'}g^{\nu\nu'} + g^{\mu\nu'}g^{\nu\mu'}-
{2\over D}
g^{\mu\nu}g^{\mu'\nu'}+C g^{\mu\nu}g^{\mu'\nu'} \: .
\end{equation}

Eq.(\ref{dewittm}) is the most general ultra--local distance, invariant
under diffeomorphisms. In fact it must be a bilinear in $\delta
g_{\mu\nu}$; the metric tensor $G^{\mu\nu\mu'\nu'}(x,y)$ must have
support in $x=y$ and should be formed only by the $ g_{\mu\nu}$
excluding its derivatives. Introducing derivatives in
$G^{\mu\nu\mu'\nu'}$ would give a dynamical role to the measure for
the field $ g_{\mu\nu}$. The analogous metric for Euclidean
Yang-Mills theory is
\begin{equation}
(\delta A,\delta A) = \int d^D x \; \mbox{Tr}(\delta
A_{\mu}(x) \delta A_{\mu}(x)).
\end{equation}
Metric (\ref{dewittm}) will be requested to be positive definite,
and this requirement puts a restriction on $C$.

In fact after writing $\delta g_{\mu\nu}=\delta g^T_{\mu\nu}+
{g_{\mu\nu}\over D} \delta g^\lambda_{\lambda}$,
being $\delta g^T_{\mu\nu}$ the traceless part, we have
\begin{equation}
(\delta g,\delta g)= 2\int\sqrt{g}\, d^D x \: \delta g^T_{\mu\nu}
g^{\mu\mu'} g^{\nu\nu'}\delta g^T_{\mu'\nu'} +
C  \int\sqrt{g} \, d^D x \: \delta g^\lambda_\lambda
\delta g^\rho_\rho
\end{equation}
from which we see that $C>0$ if we want a positive definite metric.
The next problem is to factor out from ${\cal D}[g]$ the
infinite volume of the diffeomorphisms and leave an integral on the
$dl_i$ multiplied by a proper Jacobian; the calculation of such a
Jacobian is the most relevant part in the process of the reduction of
the integral on the metrics to the integral over the geometries.

We have to generalize to an infinite dimensional space the usual
procedure which relates a distance (metric) to a volume element
(measure). We stress that even in the case when our parameters $l_i$
are finite in number, the integration space on the metric is always
infinite dimensional due to the presence of the diffeomorphisms
$f$. In a finite dimensional space $t_1, t_2, \dots , t_n$ with
distance $(\delta t,\delta t)= \delta t_i M^{ij}(t)\delta t_j $ the
integration measure is given by
\begin{equation}
J(t) \prod_i dt_i = \sqrt{\det M(t)} \prod_i dt_i
\end{equation}
and such $J(t)$ can be computed by means of an integration on the
 tangent space at the point $t$ i.e.
\begin{equation}
1= {J(t)\over (2\pi)^{N/2}}\int  \prod_i d\delta t_i \;  e^{-{1\over 2} 
(\delta t,\delta t)}.
\end{equation}

Similarly one proceeds on the infinite dimensional space generated by
the diffeomorphisms i.e.  $g_{\mu\nu}(x,l,f)= [f^{\star}\bar
g_{\mu\nu}(l)](x)$ by writing, apart from an irrelevant multiplicative
constant
\begin{equation}
\label{functionalm}
1= \int {\cal D}[\delta g] e^{-{1\over 2} (\delta g,\delta g)}= 
J(l,f) \int\prod_i d\delta l_i {\cal D}[\xi]
e^{-{1\over 2} (\delta g,\delta g)}
\end{equation}
where
\begin{equation}
\delta g_{\mu\nu}(x) = f^{\star}[\overline{\nabla}_{\mu}
\bar{\xi}_{\nu} +\overline{\nabla}_{\nu} \bar{\xi}_{\mu} ](x) + \left[
  f^{\star} {\partial \bar g_{\mu\nu}(l) \over dl_i} \delta l_i
\right](x)  = 
\nabla_\mu \xi_{\nu}(x) +\nabla_\nu \xi{_\mu}(x) + {\partial 
g_{\mu\nu}(x,l,f) \over \partial l_i} \delta l_i \: .
\end{equation}
The first term in the variation can be understood as $g_{\mu\nu}(x, l,
f_{1} \cdot f) - g_{\mu\nu}(x,l,f)$, where $f_{1}$ is the
infinitesimal diffeomorphism $x^{\mu} \rightarrow x^{\mu} +
\bar{\xi}^{\mu} (x)$; $\nabla_\mu$ is the covariant derivative in the
metric $[f^{\star} \bar g_{\mu\nu}(l)](x) = g_{\mu\nu}(x,l,f)$.  In
eq.(\ref{functionalm}) ${\cal D}[\xi]$ is defined by adopting,
analogously to eq.(\ref{dewittm}) the diffeomorphism invariant
distance
\begin{equation}
\label{fmetric}
(\xi, \xi) = \int \sqrt{g(x)}\, d^D x \,\xi_{\mu}
(x) g^{\mu\nu}(x) \xi_{\nu} (x)
\end{equation}
i.e.
\begin{equation}
1= \int {\cal D}[\xi] \: e^{-{1\over 2}(\xi, \xi)}.
\end{equation}
Eq.(\ref{fmetric}) defines, analogously to what happens in Yang--Mills
theory for the gauge transformations, the ultra--local distance
between two diffeomorphisms. As a result $J$ appearing in
eq.(\ref{functionalm}) is independent of $f$.  To compute $J$ we shall
need to decompose $\delta g_{\mu\nu}$ in a part orthogonal to the
gauge orbits generated by the diffeomorphisms and a remainder. In
order to do this we have to discuss in more detail the operator $F$
defined by $(F\xi)_{\mu\nu} = \nabla_{\mu} \xi_{\nu} + \nabla_{\nu}
\xi_{\mu}$. $F^\dagger$ is the adjoint of $F$ according the positive
definite metric (\ref{dewittm}). $F$ acts on the vector fields
$\xi_{\mu}$ whose Hilbert space, equipped with the norm provided by
the Lebesgue measure $L^2(\sqrt{g}\, d^Dx \,g^{\mu\nu})$, will be
denoted by $\Xi$. The result of $F$ acting on $\Xi$ are symmetric
tensor fields. We shall denote by ${\cal H}$ the Hilbert space of the
symmetric tensor fields $h$ equipped with the norm provided by the
Lebesgue measure $L^2(\sqrt{g}\, d^Dx\, G^{\mu\nu\mu'\nu'})$. It is
well known that ${\cal H}$ can be decomposed as ${\cal H} = \overline{
\mbox{Im}(F)}\oplus \mbox{Ker}(F^\dagger)$ and $\Xi$ as $\Xi =
\overline{ \mbox{Im}(F^\dagger)}\oplus \mbox{Ker}(F)$. In physical
terms $\mbox{Ker} (F)$ represents the Killing vector fields of the
metric (if they exist) while $\mbox{Ker} (F^\dagger) $ corresponds to
the true variations of the geometry. On a finite dimensional space the
operator $F^\dagger F$ acting on $\mbox{Im} F^\dagger$ onto $\mbox{Im}
F^\dagger$ has a well defined inverse. With infinite dimensions, as it
is our case, some assumption is needed. Let us consider the equation
\begin{equation}
\label{inversion}
F^\dagger F \eta = F^\dagger h
\end{equation} 
with $h_{\mu\nu}={\partial g_{\mu\nu}\over \partial l_i}$. We can
decompose $h$ as $h=h_0+h_1$ with $h_0\in \mbox{Ker}(F^\dagger)$ and
$h_1\in \overline{\mbox{Im}(F)}$. The regularity assumption will be
that the family $\bar{g}_{\mu\nu}(x,l)$ has been chosen so that $h\in
D(F^\dagger)$ and $h_1\in \mbox{Im}(F)$. Physically it means that the
``gauge part'' of $h$ can be written as $\nabla_\mu \xi_\nu+\nabla_\nu
\xi_\mu = (F\xi)_{\mu\nu} $ and not as a singular limit of gauge
transformations. Then it is immediate that the solution of
eq.(\ref{inversion}) in $\overline{\mbox{Im}(F^\dagger)}$ is given by
$\eta=\xi_1$ where $\xi=\xi_0+\xi_1$ with $\xi_0\in \mbox{Ker}(F)$ and
$\xi_1\in \overline{\mbox{Im}(F^\dagger)}$. Due to the positive
definite metric (\ref{fmetric}) such solution is unique.

We can now decompose the variation of the metric into two orthogonal
parts as follows  
\begin{equation}
\label{decomp}
\delta g_{\mu\nu} = [ (F\xi)_{\mu\nu} + F(F^\dagger F)^{-1} 
F^\dagger {\partial g_{\mu\nu}\over \partial l_i} \delta l_i ] +
[1 - F (F^\dagger F)^{-1} F^\dagger] 
{\partial g_{\mu\nu}\over \partial l_i} \delta l_i
\end{equation}
and $(F^\dagger F)^{-1} F^\dagger {\partial g_{\mu\nu}\over \partial
l_i} \delta l_i$ can be absorbed in a shift of $\xi_\mu$.  Obviously
the optimal choice for $\bar g_{\mu\nu}(x,l)$ would be such that
\begin{equation}
\label{ortho}
{\partial \bar g_{\mu\nu}\over \partial l_i}\in \mbox{Ker}(\bar{F}^\dagger)
\end{equation}
in which case the two terms $(F\xi)_{\mu\nu}$ and ${\partial
g_{\mu\nu}\over \partial l_i}$ would be already orthogonal, saving the
effort to compute the inverse of $F^\dagger F$ on
$\overline{\mbox{Im}(F^\dagger)}$.

However in general this choice cannot be accomplished. In fact we show
in Appendix A that for a generic choice of geometries described by the
parameters $l$ there is no related gauge fixing surface which is
orthogonal to the gauge fibers.

On the other hand if the class of geometries described by the $l$ are
such that eq.(\ref{ortho}) is satisfied, then such a property
holds all along the gauge fiber $f^{\star} \bar{g}_{\mu\nu}(l)$.

Substituting eq.(\ref{decomp}) into eq.(\ref{functionalm}) we have
\begin{equation}
\label{jacob}
J(l) =  \det (t^i, t^j)^{\frac{1}{2}} {\cal D}\mbox{et} 
(F^\dagger F)^{\frac{1}{2}}
\label{see}
\end{equation}
with
\begin{equation}
t^i_{\mu\nu}=[1 - F (F^\dagger F)^{-1} F^\dagger] 
{\partial g_{\mu\nu}\over \partial l_i}.
\end{equation}
${\cal D}\mbox{et} (F^\dagger F)$ is a true functional determinant and
it is the Faddeev--Popov corresponding to the gauge fixing $\bar
g_{\mu\nu}(l)$. We notice in this connection that, provided the
parameters $l$ are geometric invariants, no Gribov problem arises, as
a diffeomorphism cannot connect two different geometries.  As $F$ is a
covariant operator the value of ${\cal D}\mbox{et} (F^\dagger F)$ does
not depend on the diffeomorphism $f$ i.e. ${\cal D}\mbox{et}
(F^\dagger F)= {\cal D}\mbox{et} (\bar F^\dagger \bar F)$, being $\bar
F$ the operator computed in the metric $\bar{g}_{\mu\nu}(l) $. The
same invariance property holds for $\det(t^i, t^j)$ with the result
that $J$ does not depend on $f$. We remark that both determinants in
eq.(\ref{jacob}) depend on the parameter $C$ appearing in the De Witt
supermetric. Such a dependence in general does not cancel out (see
Appendix B). Actually the dependence on $C$ could also be taken as a
index of the approach to the continuum theory when the number of the
parameters $l$ becomes large. In the next section we shall see how the
integration over the conformal factor makes the result
$C$-independent.

As we pointed out at the beginning of this section, the De Witt metric
eq.(\ref{dewittm}) is a diffeomorphism invariant provided the transition
functions are independent of the metric. In fact only in this case
$\delta g_{\mu\nu}$ transforms like a tensor.  

Having reached the two expressions ${\cal D}\mbox{et} (F^\dagger F)$
and $\det(t^i, t^j)$ which are invariant under {\it
rigid} diffeomorphisms, it is of interest to consider $l$--dependent
diffeomorphisms, which modifies the gauge fixing surface in an
$l$--dependent way. As the Faddeev--Popov \ term ${\cal D}\mbox{et}
(F^\dagger F)$ does not depend on any derivative of $g_{\mu\nu}$ with
respect to $l$, it is left invariant under $l$--dependent
diffeomorphisms. With regard to $\det(t^i, t^j)$ we
first examine the behavior of $\frac{\partial \bar{g}_{\mu\nu}
(x,l)}{\partial l_{i}}$ under such diffeomorphisms
\begin{equation}
\bar{g}_{\mu\nu}(x,l) = A^{\lambda'}_{\mu} A^{\rho'}_{\nu}
\bar{g'}_{\lambda'\rho'}(x'(x,l),l)  
\end{equation}
with $\displaystyle A^{\lambda'}_{\mu} = \frac{\partial {x'}^{\lambda'}
  (x,l)}{\partial x^{\mu}}$. Consequently 
\begin{eqnarray}
\lefteqn{ \frac{\partial \bar{g}_{\mu\nu} (x,l)}{\partial l_{i}} =
  A^{\lambda'}_{\mu} A^{\rho'}_{\nu} \frac{\partial
  {\bar{g}'}_{\lambda'\rho'} 
    (x'(x,l),l)}{\partial l_{i}} } & & \\
& & + A^{\lambda'}_{\mu} A^{\rho'}_{\nu} 
\frac{\partial {x'}^{\alpha}(x,l)}{\partial l_{i}} 
\frac{\partial {\bar{g}'}_{\lambda'\rho'}(x'(x,l),l)}{\partial
  {x'}^{\alpha}} 
+ 2 \frac{\partial  A^{\lambda'}_{\mu}}{\partial l_{i}}
  A^{\rho'}_{\nu} {\bar{g}'}_{\lambda'\rho'}(x'(x,l),l) \: ,
\nonumber
\end{eqnarray}
which shows that $\frac{\partial \bar{g}_{\mu\nu} (x,l)}{\partial l_{i}}$
is not a tensor under this class of transformations.

Setting
\begin{equation}
\bar{F}(\bar{F}^{\dag}\bar{F})^{-1} \bar{F}^{\dag} \frac{\partial
\bar{g}_{\mu \nu}}{\partial l_{i}}= \bar{B}^{\lambda\rho}_{\mu\nu}
\frac{\partial \bar{g}_{\lambda \rho}}{\partial l_{i}}
\end{equation}
the second term in eq.(\ref{decomp}) transforms as 
\begin{eqnarray}
\lefteqn{ (\delta^{\lambda}_{\mu} \delta^{\rho}_{\nu} -
  \bar{B}^{\lambda \rho}_{\mu 
  \nu})  \frac{\partial \bar{g}_{\lambda \rho}}{\partial l_{i}} =
A^{\mu'}_{\mu} A^{\nu'}_{\nu} (\delta^{\lambda'}_{\mu'}
\delta^{\rho'}_{\nu'} - {\bar{B'}}^{\lambda' \rho'}_{\mu' \nu'})
(A^{-1})^{\alpha}_{\lambda'} (A^{-1})^{\beta}_{\rho'} \cdot}  \\
&& \cdot \left[ A^{\alpha'}_{\alpha} A^{\beta'}_{\beta}  \frac{\partial
    {\bar{g}'}_{\alpha' \beta'}}{\partial l_{i}} + A^{\alpha'}_{\alpha}
  A^{\beta'}_{\beta}  \frac{\partial {x'}^{\gamma'} }{\partial l_{i}}
  \frac{\partial {\bar{g}'}_{\alpha' \beta'}}{\partial {x'}^{\gamma'}}  
+ 2 \frac{\partial A^{\alpha'}_{\alpha}}{\partial l_{i}}
A^{\beta'}_{\beta} {\bar{g}'}_{\alpha' \beta'} \right] = \nonumber \\ 
&& A^{\mu'}_{\mu} A^{\nu'}_{\nu} (\delta^{\lambda'}_{\mu'}
\delta^{\rho'}_{\nu'} - {\bar{B'}}^{\lambda' \rho'}_{\mu' \nu'}) \left[
 \frac{\partial {\bar{g}'}_{\lambda' \rho'}}{ \partial l_{i}} + 
\frac{\partial {x'}^{\gamma'} }{\partial l_{i}}
  \frac{\partial {\bar{g}'}_{\lambda' \rho'}}{\partial {x'}^{\gamma'}}
  +  2 (A^{-1})^{\alpha}_{\lambda'} \frac{\partial
A^{\alpha'}_{\alpha}}{\partial l_{i}} {\bar{g}'}_{\alpha' \rho'} \right]
= \nonumber  \\
&& A^{\mu'}_{\mu} A^{\nu'}_{\nu} (\delta^{\lambda'}_{\mu'}
\delta^{\rho'}_{\nu'} - {\bar{B'}}^{\lambda' \rho'}_{\mu' \nu'}) \left[ 
\frac{\partial {\bar{g}'}_{\lambda' \rho'}}{ \partial l_{i}} +
{\overline{\nabla}'}_{\lambda'} \xi_{\rho'} +
  {\overline{\nabla}'}_{\rho'} \xi_{\lambda'} 
\right] \nonumber
\end{eqnarray}
where $\displaystyle \xi^{\rho'} =
\frac{\partial{x'}^{\rho'}(x,l)}{\partial l_{i}}$. But then the
projector $(\delta^{\lambda'}_{\mu'} \delta^{\rho'}_{\nu'} -
{\bar{B'}}^{\lambda' \rho'}_{\mu' \nu'})$ annihilates the
$\bar{F}'\xi$ part and we are left with
\begin{equation}
A^{\mu'}_{\mu} A^{\nu'}_{\nu} (\delta^{\lambda'}_{\mu'}
\delta^{\rho'}_{\nu'} - {\bar{B'}}^{\lambda' \rho'}_{\mu' \nu'})
\frac{\partial {\bar{g}'}_{\lambda' \rho'}}{\partial l_{i}} 
\end{equation}
i.e.\ the same covariant expression as under a rigid
diffeomorphism. We stress that this invariance is due to the
appearance of the projector $(I - \bar{B})$. The larger freedom on the
diffeomorphism transformations may be useful in the difficult job of
computing the functional determinant of the Lichnerowicz operator on
the manifold. Physically we found, starting from the De Witt
supermetric, a geometric invariant measure which depends only on the
geometries given by the $l_{i}$ and not on the particular metric used
in describing them. We notice finally that a change of the parameters,
i.e.\ $l_{i} \longrightarrow {l'}_{i} (l)$ leaves the result invariant
in form.

\section{Measure for the conformal factor}
\label{tre}

We pointed out in the previous section that the integration on the
parameters $l$ leaves a dependence of the result on the constant $C$
appearing in the De Witt supermetric. In this section we want to
enlarge the treatment replacing the integration variables $l_{i}$ by a
conformal factor $\sigma(x)$ \cite{mazm} and a finite number of other
parameters $\tau_{i}$ describing geometric deformations transverse
(i.e.\ non collinear) both to the diffeomorphism and to the Weyl
group.

Thus the set of metrics we shall integrate on is given by
\begin{equation}
  g_{\mu\nu} (x,\tau,\sigma,f) = [f^{\star} e^{2\sigma} \hat{g}_{\mu\nu}
  (\tau) ](x) \; .
\nonumber
\end{equation}
In the following we denote by $\bar{g}_{\mu\nu} (x,\tau,\sigma)$ the
combination
\begin{equation}
  \bar{g}_{\mu\nu} (x,\tau,\sigma) = e^{2\sigma} \hat{g}_{\mu\nu} (x,\tau).
\nonumber
\end{equation}
We have to evaluate the Jacobian $J(\sigma, \tau)$ such that
\begin{equation}
  {\cal D}[g] = J(\sigma, \tau) {\cal D}[f] {\cal D}[\sigma]
  \prod_{i} d\tau_{i} \; . 
\label{triangolo}
\end{equation}
We proceed as in sect.2. The general variation of the metric can be
written as \cite{allc}
\begin{equation}
  \delta g_{\mu\nu} (x,\tau,\sigma,f) = (F \xi)_{\mu\nu}(x) + 2
  [f^{\star} \delta \sigma \bar{g}_{\mu\nu}](x) +
  [f^{\star} \frac{\partial \bar{g}_{\mu\nu}}{\partial \tau_{i}}
  \delta \tau_{i}](x) \; .  
\end{equation}
Defining the operator $P$ by
\begin{eqnarray}
\lefteqn{(P\xi)_{\mu\nu} = (F\xi)_{\mu\nu} - \frac{g_{\mu\nu}}{D}
g^{\alpha \beta} (F \xi)_{\alpha \beta} =} && \\
&& = \nabla_{\mu} \xi_{\nu} + \nabla_{\nu} \xi_{\mu} -
\frac{2}{D} g_{\mu\nu} \nabla \cdot \xi \nonumber
\end{eqnarray}
and the traceless tensor
\begin{equation}
k^{i}_{\mu\nu} = \frac{\partial g_{\mu\nu}}{\partial \tau_{i}} -
\frac{g_{\mu\nu}}{D} g^{\alpha\beta} \frac{\partial g_{\alpha
      \beta}}{\partial \tau_{i}} 
\end{equation}
we can rewrite
\begin{equation}
  \delta g_{\mu\nu} (x,\tau,\sigma,f) = (P\xi)_{\mu\nu}(x) + f^{\star}
  \left[ \left( 2\delta \sigma + \frac{\bar{g}^{\alpha\beta}}{D}
  \frac{\partial 
  \bar{g}_{\alpha\beta}}{\partial \tau_{i}} \delta \tau_{i}  +
    \frac{\bar{g}^{\alpha\beta}}{D} (\bar{F}\bar{\xi})_{\alpha \beta}
  \right) \bar{g}_{\mu\nu}(\sigma,\tau) \right](x) + k^{i}_{\mu\nu}(x)
  \delta 
\tau_{i}. 
\end{equation}
Setting now 
\begin{equation}
  \xi'_{\mu} = \xi_{\mu} + \left( \frac{1}{P^{\dag} P} P^{\dag}
  k^{i} \delta \tau_{i} \right)_{\mu}
\end{equation}
and 
\begin{equation}
  \delta \sigma' = \delta \sigma + \frac{\bar{g}^{\alpha\beta}}{2D}
      \frac{\partial \bar{g}_{\alpha\beta}}{\partial \tau_{i}} \delta
      \tau_{i} + \frac{\bar{g}^{\alpha\beta}}{2D} (\bar{F}\bar \xi
      )_{\alpha \beta}
\label{shifts}
\end{equation}
we obtain 
\begin{equation}
  \delta g_{\mu\nu} (x,\tau,\sigma,f) = (P\xi')_{\mu\nu} + f^{\star}
  2\delta \sigma' \bar{g}_{\mu\nu} (\tau,\sigma)  + \left( 1 - P
    \frac{1}{P^{\dag} P} P^{\dag} \right) k^{i}_{\mu\nu} \delta \tau_{i}.
\end{equation}
We remark that the three terms are mutually orthogonal and thus
\begin{eqnarray}
\lefteqn{1 = \int {\cal D} [\delta g] e^{-\frac{1}{2} (\delta
    g ,\delta g ) } = }  & & \\
&& J(\sigma, \tau) \int {\cal D} [\delta \sigma ] {\cal D} [\xi]
   \prod_{i} d\tau_{i} e^{-\frac{1}{2} (P \xi' , P \xi'
    )} \cdot e^{-2 CD^{2} (\delta \sigma', \delta
   \sigma')} \cdot \nonumber \\
&& \cdot e^{-\frac{1}{2} (( 1 - \bar{P} \frac{1}{\bar{P}^{\dag} \bar{P}}
   \bar{P}^{\dag}) \bar{k}^{i} \delta \tau_{i}  , ( 1 - \bar{P}
   \frac{1}{\bar{P}^{\dag} \bar{P}} \bar{P}^{\dag})
   \bar{k}^{j} \delta \tau_{j}  )} \; . \nonumber
\end{eqnarray}

Exploiting invariance under translations of the integrals on the
tangent space and the definition of ${\cal D} [\delta \sigma ]$
\begin{equation}
  \int {\cal D} [\delta \sigma ] e^{-\frac{1}{2} (\delta \sigma, \delta
   \sigma)}=1
\end{equation}
with $(\delta \sigma, \delta \sigma)= \int \sqrt{g(x)} d^{D} x \delta
\sigma(x) \delta \sigma(x)$ we have apart for a multiplicative constant
\begin{equation}
\displaystyle
  J(\sigma, \tau) = {\cal D}\mbox{et} (\bar{P}^{\dag} \bar{P}
  )^{\frac{1}{2}} 
    \left[ \det \left(  \bar{k}^{i}, ( 1 - \bar{P}
   \frac{1}{\bar{P}^{\dag} \bar{P}} \bar{P}^{\dag})
   \bar{k}^{j} \right) \right]^{\frac{1}{2}} \: . 
\label{quadrato}
\end{equation}

The dependence on $f$ has disappeared due to the invariance of the De
Witt metric under diffeomorphisms and thus in eq.(\ref{triangolo}) the
infinite volume of the diffeomorphisms can be factorized away.

We notice that in eq.(\ref{quadrato}) the dependence on $C$ has been
absorbed in an irrelevant multiplicative constant, as it happens in two
dimensions. This is the result of having integrated over all the
conformal deformations.  On the other hand the $\bar
k^i_{\mu\nu}$, depend both on $\tau$ and $\sigma$ and also the operators
$\bar P$, $\bar P^\dagger$ depend on $\tau$ and $\sigma$ through the
metric $\bar g_{\mu\nu}$
\begin{equation} 
\bar P = e^{2\sigma} \hat P e^{-2\sigma}, \mbox{\ \ \ \ } \bar
P^\dagger=e^{-D\sigma} \hat P^\dagger e^{(D-2)\sigma}, \mbox{\ \ \ \ }
\bar 
P^\dagger \bar P = e^{-D\sigma} \hat P^\dagger e^{D\sigma} \hat P
e^{-2\sigma}. 
\end{equation} 
For the subsequent discussion, it will be useful to examine
$\mbox{Ker}(\bar P)$ and $\mbox{Ker}(\bar P^\dagger )$. As already
mentioned, $\mbox{Ker}(\bar P)$ is given by the conformal Killing
vectors of the geometry $\bar g_{\mu\nu}(x,\tau,\sigma)$. In fact
under the infinitesimal diffeomorphism generated by $\bar{\xi}$
\begin{eqnarray} 
\bar g_{\mu\nu} & \rightarrow & \bar g_{\mu\nu} + (\overline
\nabla_\mu \bar{\xi}_\nu +\overline \nabla_\nu \bar{\xi}_\mu -{2\over
  D} \bar 
g_{\mu\nu} \overline \nabla\cdot \bar{\xi})+ {2\over D} \bar
g_{\mu\nu}\overline \nabla\cdot \bar{\xi} = \nonumber \\
& = & \bar g_{\mu\nu} (1+{2\over D}\overline \nabla\cdot \bar{\xi})
+(\bar P 
\bar{\xi})_{\mu\nu} = \bar g_{\mu\nu} (1+{2\over D}\overline
\nabla\cdot \bar{\xi}) 
\end{eqnarray} 
if $(\bar P \bar{\xi})_{\mu\nu}=0$. Contrary to what happens in two
dimensions, where every topology carries its own conformal Killing
vectors (6 for the sphere, 2 for the torus and 0 for higher genus),
here we shall have no conformal Killing vectors for a generic $\hat
g_{\mu\nu}(\tau)$, and thus the geometries with $\mbox{Ker} \bar P
\neq \emptyset$ have zero relative measure. The null eigenvectors
$\bar{\xi}$ of $\bar P$ are related to those of $\hat P$ by $\bar{\xi}
= e^{2\sigma}\hat \xi$ where $\hat\xi$ are independent of
$\sigma$. Similarly the null eigenvectors $\bar{h}$ of $\bar
P^\dagger$ are related to those of $\hat P^\dagger$ by $\bar h=
e^{(2-D)\sigma}\hat h$. The $\mbox{Ker}(\bar P^{\dagger})$ is the
analogous of the pure Teichm\"uller deformations in two dimensions. On
the other hand we have
\begin{equation}
\bar k^i = e^{2\sigma}\hat k^j \; .
\label{sigmadep}
\end{equation}
We notice that given an orthonormal basis $\hat h^n$ of $\mbox{Ker}(\hat
P^\dagger)$, the vectors $\bar{h}^n= e^{(2-D)\sigma}\hat h^n$, even
though complete in $\mbox{Ker}({\bar P}^{\dagger})$ do
not remain orthogonal as
\begin{equation}
(\bar{h}^m,\bar{h}^n) = 2\int \sqrt{\hat g}\,d^D x\, e^{-D\sigma}
\hat h^m_{\mu\nu} \hat
g^{\mu\mu'}\hat g^{\nu\nu'} \hat h^n_{\mu'\nu'}.
\end{equation}
On the other hand from  eq.(\ref{sigmadep})
\begin{equation}
(\bar h^m, \bar k^i) = (\hat h^m,\hat k^i).
\label{confinv}
\end{equation}
The operator  $1-\bar P{1\over \bar P^\dagger \bar P}\bar P^\dagger$
appearing in eq.(\ref{quadrato}) projects on $\mbox{Ker}(\bar
P^\dagger)$ and thus can be written in terms of the $\bar h^n_{\mu\nu}$ as 
\begin{equation}
\left( 1-\bar P{1\over \bar P^\dagger \bar P}\bar P^\dagger
\right)_{\mu\nu, \lambda\rho}
(x,x') = \bar h^m_{\mu\nu} (x) \bar M^{-1}_{mn} \bar h^n_{\lambda\rho}
(x') 
\end{equation}
where $M_{mn}$ is the infinite matrix
\begin{equation}
M_{mn} = (h^m, h^n). 
\end{equation}
The $( 1-\bar P{1\over \bar P^\dagger \bar P}\bar P^\dagger ) \bar
k^{i}$ span an $N$--dimensional subspace of $\mbox{Ker} (\bar
P^{\dagger})$. In fact if $a_{i}$ exist such that $( 1-\bar P{1\over
\bar P^\dagger \bar P}\bar P^\dagger ) \sum_{i=1}^{N} a_{i} \bar k^{i}
=0 $ then it would mean that $\sum_{i=1}^{N} a_{i} \bar k^{i} = \bar P
\xi$.  Thus $\sum_{i=1}^{N} a_{i} \frac{\partial \bar
g_{\mu\nu}}{\partial \tau_{i}}$ would be the sum of a Weyl and a
diffeomorphism transformation.

For $\sigma=0$ we can choose the $h^n$ with the properties $(h^n,
k^i)=0$ for $n > N$ and a non zero determinant of the $N\times N$
matrix $(h^n, k^i)$ with $i,n \leq N$. Such properties are maintained
for $\sigma \neq 0$, due to the independence on $\sigma$ in
eq.\ref{confinv}, but all matrix elements of $M_{mn}$ are needed to
compute the top left $N\times N$ sub-matrix of $M^{-1}_{mn}$ whose
determinant we shall denote by $\det M^{-1}_{N\times N}$. The
computation of such a sub-matrix and of the Faddeev--Popov determinant
${\cal D}\mbox{et}(\bar P^\dagger \bar P)$ are the two technically
difficult points.

We now examine the functional dependence on $\sigma$ of the two
determinants. We notice that
\begin{eqnarray}
\det \left( (1-\bar P{1\over \bar P^\dagger \bar P}\bar
P^\dagger)\bar k^i, (1-\bar P{1\over \bar P^\dagger \bar P}\bar
P^\dagger)\bar k^j \right) & = &
[\det(\bar k^i, \bar h^n)]^2 \det \bar M^{-1}_{N\times N}
\nonumber \\ & = &[\det(\hat k^i, \hat h^n)]^2 \det
\bar M^{-1}_{N\times N} \nonumber \\
\end{eqnarray}
due to eq.(\ref{confinv}). Thus the dependence on $\sigma$ is
restricted to ${\cal D}\mbox{et}(\bar P^\dagger \bar P)$ and to $\det
\bar M^{-1}_{N\times N}$. We want to stress at this point the main
differences between $D=2$ and $D>2$. In $D=2$, $\mbox{Ker}(P^\dagger)$
is finite dimensional (quadratic differentials), and thus the number
of the $k^i$ is finite and $(1 - P (P^{\dag}P)^{-1} P^{\dag})k^{i}$
span completely $\mbox{Ker}(P^\dagger)$.  As for ${\cal
D}\mbox{et}(\bar P^\dagger \bar P)$ its dependence on $\sigma$ in
$D=2$ is obtained by computing the variation under $\delta \sigma$ and
then integrating back as a result one obtains the Liouville action. In
$D>2$ (referring to the generic case in which there are no conformal
Killing vectors) from
\begin{equation}
\log {\cal D}\mbox{et} (\bar{P}^{\dag} \bar{P}) = - \frac{d}{ds} Z(0) = 
- \frac{d}{ds} \left[ \frac{1}{\Gamma (s)} \int_{0}^{\infty} \! dt \:
t^{s -1} \mbox{Tr} (e^{-t \bar{P}^{\dag}\bar{P}}) \right]
\end{equation}
we have
\begin{equation}
-\delta \log{\cal D}\mbox{et} (\bar P^\dagger \bar P)
= \gamma_E \delta Z_{\bar P^\dagger \bar P}(0) + 
{\mbox{Finite}}_{\epsilon \rightarrow 0}
\int^\infty_\epsilon dt \{(2+D)\mbox{Tr}(e^{-t \bar P^\dagger \bar P}
\delta\sigma) -D  \mbox{Tr}'(e^{-t \bar P\bar P^\dagger}\delta\sigma)\},
\end{equation}
where $\mbox{Tr}'$ excludes the $0$-modes of $\bar P^\dagger$, which
now ($D>2$) are infinite in number.
We notice that ${\cal D}\mbox{et}(\bar P^\dagger \bar P)$ is well
defined because  $\bar P^\dagger \bar P$ is an elliptic operator
\cite{ell}. 

In fact 
\begin{equation}
P^{\dag} P \xi_{\nu} = -4 [ \nabla^{2} \xi_{\nu} + \nabla^{\mu}
\nabla_{\nu} \xi_{\mu} - \frac{2}{D} \nabla_{\nu} \nabla \cdot \xi ]
\end{equation}
and the determinant of the leading symbol \cite{ell}
\begin{equation}
4[ k^{2} \delta_{\mu}^{\mu'} + ( 1 - \frac{2}{D} ) k_{\mu} k^{\mu'} ]
\end{equation}
vanishes only for $k=0$. On the other hand
\begin{equation}
PP^{\dag} h_{\mu\nu} = -4 (\nabla_{\nu} \nabla^{\lambda}
h_{\lambda\mu} + \nabla_{\mu} \nabla^{\lambda} h_{\lambda\nu}  -
\frac{2}{D} g_{\mu\nu}  \nabla^{\rho} \nabla^{\lambda} h_{\lambda\rho}
)
\end{equation}
whose leading symbol
\begin{equation}
2[ k_{\nu} k^{\mu'} \delta_{\mu}^{\nu'} + k_{\mu} k^{\mu'}
\delta_{\nu}^{\nu'} 
+ k_{\nu} k^{\nu'} \delta_{\mu}^{\mu'} + k_{\mu} k^{\nu'}
\delta_{\nu}^{\mu'}] 
-\frac{8}{D} \delta_{\mu\nu}  k^{\mu'} k^{\nu'} 
\end{equation}
has zero determinant for $k\neq 0$ as it is immediately seen by
applying it to a tensor of the form $h_{\mu\nu} = v_{\mu} w_{\nu}
+v_{\nu} w_{\mu}  $ 
with $w\cdot v = w\cdot k = v\cdot k = 0$. Thus the variation with
respect to $\sigma$ cannot be computed in term of local quantities as
the usual heat kernel technique is not available. An exception is the
variation with $\delta \sigma = {\rm const.}$ under which due to $
\mbox{Tr}(e^{-t \bar P^\dagger \bar P}) = \mbox{Tr}'(e^{-t \bar P\bar
P^\dagger})$ the calculation can be reduced to the heat kernel of the
elliptic operator $\bar P^\dagger \bar P$. For the expression of such
variation see \cite{anto}.

\section{Conclusions}

Summarizing, for a class of metrics of the type $f^{\star}  
\bar{g}_{\mu\nu} (l)$ the De Witt supermetric induces
unambiguously the $C$--dependent measure eq.(\ref{see})
\begin{equation}
\prod_{k} dl_{k} [ \det (t^i, t^j) {\cal D}\mbox{et} 
(F^\dagger F) ]^{\frac{1}{2}} \: .
\label{measure1}
\end{equation}
Similarly for the class $f^{\star} e^{2\sigma} \hat{g}_{\mu\nu} (\tau)$
we have the $C$--independent measure eq.(\ref{quadrato})
\begin{equation}
\displaystyle
\prod_{k} d\tau_{k} {\cal D}[\sigma] \left[  {\cal D}\mbox{et} (
  P^{\dag} P ) 
  \det \left(  k^{i}, ( 1 - P
  \frac{1}{P^{\dag} P} P^{\dag})
  k^{j} \right) \right]^{\frac{1}{2}}  
\label{measure1.99}
\end{equation}
with ${\cal D}[\sigma]$ the measure induced by the distance
\begin{equation}
(\delta \sigma, \delta \sigma)= \int d^{D} x \sqrt{\hat{g}(x)} \: \;
e^{D\sigma} \delta \sigma(x) \delta \sigma(x) \: .
\end{equation}
In the first case we have a finite dimensional integral and as such
more suitable to numerical calculations. A finite dimensional
approximation to eq.(\ref{measure1.99}) is obtained by restricting to
a family of conformal factors parameterized by a finite numbers of
parameters $s = \{ s_{i} \}$.  Thus to the family
$f^{\star} e^{2\sigma(s)} \hat{g}_{\mu\nu} (\tau) $ it is associated
the measure 
\begin{equation}
\displaystyle
\prod_{k} d\tau_{k} \prod_{i} ds_{i} \left[ \det(J^{\sigma}_{ij}) 
  \det \left( k^{i}, ( 1 - P
  \frac{1}{ P^{\dag} P } P^{\dag})
  k^{j} \right) 
  {\cal D} \mbox{et} (P^{\dag} P ) \right]^{\frac{1}{2}} \; , 
\label{measure2}
\end{equation}
where $J^{\sigma}_{ij} = \int d^{D} x \sqrt{\hat{g}}\;  e^{D\sigma}
\frac{\partial \sigma}{\partial s_{i}} \frac{\partial \sigma}{\partial
s_{j}} $. If now we denote by $l= \{ \tau_{i}, s_{j} \}$ and use the
first scheme eq.(\ref{measure1}), we get a different result (e.g.\ in
the first scheme the result depends on $C$ while in the second it does
not). The reason is that in deriving the measure (\ref{measure1.99})
$\sigma$ is a generic function, for which the shift (\ref{shifts})
is allowed. With $\sigma$ depending on a finite number of parameters
$s_{i}$ such shift will be the more accurate the higher the numbers of
the parameters. In this limit one expects complete equivalence of
the measure (\ref{measure1}) and (\ref{measure2}) for metrics of the
type $f^{\star} e^{2\sigma(s)} \hat{g}_{\mu\nu} (\tau)$. As
remarked in the sect.\ref{tre} the dependence on $C$ of the measure
(\ref{measure1}) is expected to drop out for a large number of $l_{i}$
describing the conformal deformations.

The measure (\ref{measure2}), with the modifications due to the
presence of the conformal Killing vectors, has been adopted in
\cite{pmppp} for the $D=2$ Regge case. The $\tau_{i}$ are the
Teichm\"uller parameters of the Riemann surfaces and the $s_{i}$
parameterize the positions and the angular defects of the conical
singularities of the Regge geometries. An exact calculation of ${\cal
D}\mbox{et} (P^{\dag} P)$ and an explicit form of $J^{s}_{ij}$ were
given.

In higher dimension $D\geq 3$ the route leading to the measure
(\ref{measure1}) appears more proficient as up to now we do not know
how to extract analytically the dependence of $ {\cal D}\mbox{et}
(\bar{P}^{\dag} \bar{P})$ and $ \det \left( \bar{k}^{i}, ( 1 - \bar{P}
\frac{1}{\bar{P}^{\dag} \bar{P}} \bar{P}^{\dag}) \bar{k}^{j} \right) $
on $\sigma(x)$.

\section*{Appendix A. A geometric property of gauge fixing surfaces}

In the text we remarked that whenever the gauge fixing surface
$g_{\mu\nu}(x,l)$ satisfies the orthogonality condition, i.e.\
\begin{equation}
F^{\dag} \left( \frac{\partial g_{\mu\nu}(x,l) }{\partial l_{i}}
\right) \equiv - 4 {\nabla}^{\mu} \frac{\partial {g}_{\mu\nu}(x,l)
  }{\partial l_{i}} - 2 \left( C - \frac{2}{D} \right) \partial_{\nu} 
\left[ {g}^{\alpha\beta}(x,l) \frac{\partial {g}_{\alpha\beta}(x,l)
    }{\partial l_{i}} \right]=0
\label{orto}
\end{equation}
great simplifications occur, because the inverse of
$F^{\dag} F$ is not to be computed. If 
\begin{equation}
F^{\dag} \left( \frac{\partial
g_{\mu\nu}(x,l) }{\partial l_{i}} \right) \neq 0
\end{equation}
it is natural to ask
for the existence of a family of diffeomorphisms $f(l)$ such that
\begin{equation}
{g'}_{\mu\nu}(x,l) = [f(l)^{\star} g_{\mu\nu}(l)] (x) = 
\frac{\partial {x'}^{\alpha}(x,l) }{\partial x^{\mu} }
g_{\alpha\beta}(x'(x,l),l) \frac{{x'}^{\beta}(x,l) }{\partial x^{\nu}}
\end{equation}
satisfies
\begin{equation}
F'^{\dag} \left( \frac{\partial {g'}_{\mu\nu}(x,l) }{\partial l_{i}}
\right) \equiv  - 4 {\nabla'}^{\mu} \frac{\partial {g'}_{\mu\nu}(x,l)
  }{\partial l_{i}} - 2 \left( C - \frac{2}{D} \right) \partial_{\nu} 
\left[ {g}'^{\alpha\beta}(x,l) \frac{\partial {g'}_{\alpha\beta}(x,l)
    }{\partial l_{i}} \right]=0 \; .
\label{reqorto}
\end{equation}

We shall prove that there exist  $g_{\mu\nu} (x,l)$ transverse to the
gauge fibers, violating eq.(\ref{orto}) for which no $f$ satisfying
eq.(\ref{reqorto}) can be found. In particular we shall show that if
an $f(l):x\rightarrow  x'(x,l)$ satisfying eq.(\ref{reqorto}) is
supposed to exist then the integrability condition
$\frac{\partial^{2} x'(x,l)}{\partial l_{i} \partial l_{j} } =
\frac{\partial^{2} x'(x,l) }{\partial l_{j} \partial l_{i} }$ 
leads to a contradiction.

Let us consider a metric $g_{\mu\nu}(x,l)$ on the
$D$--dimensional torus represented by an hypercube $0 \leq x^{\mu}<
1$ with opposite faces identified, with the properties
\begin{equation}
g_{\mu\nu}(x,0) = \delta_{\mu\nu} , \mbox{\ \ \ \ }
\left. \delta^{\alpha\beta} \frac{\partial g_{\alpha\beta}(x,l)
}{\partial l_{i}}\right|_{l=0}=0 , \mbox{\ \ \ \ }
\left.\delta^{\alpha\beta} \partial_{\alpha} \frac{\partial
g_{\beta\nu}(x,l) }{\partial l_{i}}\right|_{l=0}=0 \;
,\label{constrmetr}
\end{equation}
for a pair of indexes $i$ and $j$.
Suppose there exists ${x'}^{\mu} = {x'}^{\mu} (x,l)$ such that
\begin{equation}
{g'}_{\mu\nu}(x,l) = \frac{\partial {x'}^{\alpha}(x,l) }{\partial x^{\mu} }
g_{\alpha\beta}(x'(x,l),l) \frac{\partial{x'}^{\beta}(x,l) }{\partial
  x^{\nu}} 
\end{equation}
satisfies eq.(\ref{reqorto}). As any finite $l$--dependent
diffeomorphism can be written as $f(l) = f_{1}(l) \cdot f(0)$ with
$f_{1}(0)=$ identity and under an $l$--independent diffeomorphism
(like $f(0)$) eq.(\ref{orto}) is covariant, we can restrict ourselves
to transformations with ${x'}^{\mu}(x,0) = x^{\mu}$. For $l=0$
eq.(\ref{reqorto}) implies
\begin{equation}
\left. 2 \partial_{\mu} \frac{\partial {g'}_{\mu\nu}(x,l) }{\partial
    l_{i}} 
\right|_{l=0}+ \left( C - \frac{2}{D} \right) \left. \partial_{\nu}
    \left( 
\delta^{\alpha \beta} 
\frac{\partial {g'}_{\alpha\beta}(x,l) }{\partial l_{i}} \right)
\right|_{l=0}=0 
\end{equation}
i.e.\ taking into account eq.(\ref{constrmetr})
\begin{equation}
\left.  \partial_{\mu} \frac{\partial g_{\mu\nu}(x,l) }{\partial
l_{i}}\right|_{l=0}
+ \partial^{2} \xi^{i}_{\nu} + (C + 1 - \frac{2}{D} )\partial_{\nu}
\partial \cdot \xi^{i} = 0
\end{equation}
where we defined
\begin{equation}
\xi^{i\: \mu} = \left. \frac{\partial{x'}^{\mu}(x,l)}{\partial l_{i}}
\right|_{l=0}. 
\end{equation}
Thus because of eq.(\ref{constrmetr}) we have
$\partial^{2} \xi^{i}_{\nu} + (C + 1 - \frac{2}{D} ) \partial_{\nu}
\partial \cdot \xi^{i} = 0$, which implies $\partial^{2} \partial
\cdot \xi^{i} = 0$, i.e. on 
the torus $\partial \cdot \xi^{i} =$const.  Thus $\partial^{2}
\xi^{i}_{\mu} = 0$ and then $\xi^{i}_{\mu} =$const.
\begin{equation}
\left. \frac{\partial {g'}_{\mu\nu} (x,l)}{\partial l_{i}}
\right|_{l=0} = \left. \frac{\partial g_{\mu\nu} (x,l)}{\partial l_{i}}
\right|_{l=0}.
\label{penultima}
\end{equation}
Eq.(\ref{reqorto}) becomes
\begin{eqnarray}
0 = 2 g'^{\mu\mu'}(x,l)\left[ \partial_{\mu'} \frac{\partial
g'_{\mu\nu} (x,l)} {\partial l_{i}} -\Gamma'^\rho_{~\mu\mu'}(x,l)
\frac{\partial g'_{\rho\nu} (x,l)} {\partial l_{i}} -\Gamma'^
\rho_{~\nu\mu'}(x,l)
\frac{\partial g'_{\mu\rho} (x,l)} {\partial l_{i}} \right]+ \nonumber
\\
(C-{2\over D})\partial_\nu\left[ g'^{\alpha\beta}(x,l)\frac{\partial
g'_{\alpha\beta} (x,l)}{\partial l_{i}}\right]. \label{ultima}
\end{eqnarray}
We consider now the antisymmetric part in $(i,j)$ of the derivative
of eq.(\ref{ultima}) with respect to $l_j$, for $l=0$ i.e. using
eq.(\ref{penultima}) 
\begin{eqnarray}
\lefteqn{ 0 = 2 \left.{\partial g^{\mu\mu'}(x,l)\over \partial
  l_j}\right|_{l=0} \left.\partial_{\mu'}
{\partial g_{\mu\nu} (x,l)\over \partial l_{i}}\right|_{l=0}+
\partial_\rho\left( \delta^{\mu\mu'} \left.{\partial
  g_{\mu\mu'}(x,l)\over \partial 
 l_j}\right|_{l=0}\right) \left.{\partial
g_{\rho\nu}(x,l)\over \partial l_i}\right|_{l=0} - }\\
& & \!\!\!\!\!\!\! \partial_\nu \left(\left.{\partial
g_{\mu\rho}(x,l)\over \partial l_j}\right|_{l=0}\right)
\left.{\partial g_{\mu\rho} (x,l)\over \partial l_{i}}\right|_{l=0}
+({C\over 2} - {1\over D}) \partial_\nu\left[\left.{\partial
g^{\alpha\beta}(x,l)\over \partial l_j}\right|_{l=0} \left.{\partial
g_{\alpha\beta}(x,l)\over \partial l_i}\right|_{l=0}\right] 
-(i\leftrightarrow j). \nonumber
\end{eqnarray}
Taking into account that the square bracket is simmetric in $(i,j)$
and condition (\ref{constrmetr}) we have at last
\begin{eqnarray}
0 = \left.{\partial g^{\mu\mu'}(x,l)\over \partial l_j}\right|_{l=0}
\left.\partial_{\mu'} {\partial g_{\mu\nu} (x,l)\over \partial
l_{i}}\right|_{l=0} - {1\over 2} \left.{\partial g_{\mu\rho}
(x,l)\over \partial l_{i}}\right|_{l=0} \partial_\nu \left.{\partial
g_{\mu\rho}(x,l)\over \partial l_j}\right|_{l=0} -(i\leftrightarrow
j).
\label{integrability}
\end{eqnarray}
Let us now choose 
\begin{equation}
\left. {\partial g_{\mu\nu}(x,l)\over \partial l_i} \right|_{l=0} =
k_\mu 
\varepsilon_\nu+  k_\nu \varepsilon_\mu = \mbox{const.}
\label{const}
\end{equation}
and
\begin{equation}
\left. {\partial g_{\mu\nu}(x,l)\over \partial l_j} \right|_{l=0} =
(\eta_\mu 
\varepsilon_\nu+  \eta_\nu \varepsilon_\mu) \sin(k\cdot x)
\label{sin}
\end{equation}
with $k\cdot \varepsilon= \eta\cdot \varepsilon =k\cdot \eta=0$ with
$k_\mu=2\pi n_\mu$ with $n_\mu$ integers, to satisfy the torus
boundary conditions. Eqs.(\ref{const}) and (\ref{sin}) satisfy the
conditions (\ref{constrmetr}) but if we substitute in
eq.(\ref{integrability}) we find instead of zero the value $-k^2
\varepsilon^2 \eta_\nu \cos(k\cdot x)\neq 0$.

\section*{Appendix B. Dependence on the $C$ parameter}
We give here a simple example to show that if one does not integrate
on the conformal factor the dependence on $C$ in $J(l)= \det
(t^i_{\mu\nu}, t^j_{\mu\nu})^{{1\over 2}} {\cal D}\mbox{et} 
(F^\dagger F)^{{1\over 2}}$ (see eq.(\ref{see})) does not cancel out.

Let us consider a flat torus in $D$ dimensions  with metric $\bar
g_{\mu\nu} =  
{\rm diag}(l_1,l_2,\ldots,l_D)$. Being the metric constant we have
$\overline{\nabla}^\mu {\partial \bar g_{\mu\nu} \over \partial l_i}
=0$ and the integration measure becomes
\begin{equation}
J= [ \det({\partial \bar g_{\mu\nu} \over \partial l_i}, {\partial \bar
  g_{\mu\nu} 
\over \partial l_j}) {\cal D}\mbox{et}' (\bar F^\dagger \bar
  F) ]^{1\over 2} / V \; , 
\end{equation}
where the volume $V=(l_{1} \cdots l_{D})^{\frac{1}{2}} $ is due to the
presence of the $D$ Killing vectors \cite{allc}.
 We have
\begin{equation}
({\partial \bar g_{\mu\nu}\over \partial l_i}, {\partial \bar
g_{\mu\nu}\over \partial l_j}) = (l_1 l_2 \ldots l_{D}
)^{\frac{1}{2}} 
[2{\delta_{ij}\over l_i l_j} + (C-{2\over D}) {1\over l_i l_j}]
\end{equation}
whose determinant is linear in $C$ and given by 
\begin{equation}
2^{D-1} (\prod_{i=1}^{D}  l_{i} )^{\frac{D}{2} - 2} CD  
\label{polyn}
\end{equation}
On the other hand, as $R_{\mu\nu}=0$
\begin{equation}
\bar F^\dagger \bar F = 4 [ \delta d + (C + 2 - {2\over D}) d \delta]
\; .
\end{equation}
Due to $\delta d d\delta =d\delta \delta d=0$ we have that the non
zero eigenvalues of $\bar F^\dagger \bar F$ are either eigenvalues of
$4 \delta d$ or of $4 (C+2 - \frac{2}{D}) d \delta$ and
\begin{equation}
Z_{(\bar F^\dagger \bar F)}(s) = Z_{(4 \delta d)}(s) + Z_{(4(C+2 -
  {2\over D})d \delta)}(s) \;.
\end{equation}
Thus
\begin{equation}
{\cal D} \mbox{et}' (\bar F^{\dag} \bar F) = \left[16 (C+ 2 -
  \frac{2}{D}) \right]^{Z_{(d \delta)}(0)} e^{-2Z'_{(d \delta)}(0)}  
\end{equation}
and thus such a power behavior cannot be canceled by the polynomial
of eq.(\ref{polyn}).

\pagebreak

\bibliographystyle{plain}
\end{document}